# Generalized kinetic equations of dynamic system interacting with non-equilibrium medium[*]

M.D. Zviadadze [a,b], A.G. Kvirikadze [a],

[a]*Ivane Javakhishvili Tbilisi State University 3, Chavchavadze Ave. Tbilisi, 0128, Georgia*

*Andronikashvili Institute of Physics, 6, Tamarashvili str, 0177, Tbilisi, Georgia,*

[b]*Georgian Technical University, 77, Kostava str, Tbilisi, 0175, Georgia*

mdzviadadze@gmail.com, m.zviadadze@mail.ru

**Abstract**

The behavior of dynamical system interacting with non-equilibrium medium is investigated. Formally exact kinetic equations are derived for the statistical operator of the dynamical system and the macroscopic parameters of the medium. In the second order of perturbation with respect of interaction of the system with the medium, expression for the scattering integral is presented. The simplest applications of the obtained equations are considered.

**Key words:** non-equilibrium statistical operator, dynamical system, kinetic equations.

## 1. Introduction

The presentation of the macroscopic system in the form of a subsystem aggregate is connected usually with the possibility of the division of its freedom degrees into separate groups comparatively weakly interacting. Separation of the oscillatory and spin freedom degrees in solids and liquids or consideration of part of the system freedom degrees as an external medium can serve as an example of such presentation.

The quantum-statistical investigation of non-equilibrium processes in the interacting subsystems began to develop especially intensively decades because of the successes of the general theory of irreversible processes [1, 2].

In this paper we study the case when a system can be presented in the form of two interacting subsystems, one of which, called the dynamic system, is described exactly by means of the statistical operator $\rho_1(t)$, while the second, called the medium, is described coarsely by means of the macroscopic parameters[1]. The exact kinetic equations of the Markovian type for the statistical operator $\rho_1(t)$ and macroscopic parameters are derived and a simplified version of these equations in the second order of the perturbation theory over interaction between subsystems is brought. The evolution of the quantum system under the interaction with the non-equilibrium medium, the motion of the impurity particle in the equilibrium liquid and the behavior of the spin-system in a thermostat are considered as simple examples of the application of the obtained equations.

---

[1][*]Initial version of this article was published in proceedings of Tbilisi State University (Georgia) V 271, 137-162, 1987.

The behavior of the dynamic system in the equilibrium medium came into attention of researchers a long time ago [3]. This problem was comprehensively considered in works [5-7].

## 2. Derivation of the master equations

Let us consider the system with the Hamiltonian

$$H = H_0 + V; \quad H_0 = H_1 + H_2,$$

where $H_1$ is the Hamiltonian of the dynamic system; $H_2$ is the basic Hamiltonian of the medium; $V$ includes the interaction between the subsystems and also those interactions inside the medium that are relatively small and do not influence the choice of the macroscopic parameters. In accordance with the idea about the reduced description of the non-equilibrium states [6], we suppose that, for times $t \gg \tau_0$ ($\tau_0$ is the chaotization time), the system state is determined by the coursed statistical operator $\rho(t)$, which depends on time implicitly, by means of the statistical operator $\rho_1(t)$ of the dynamic system and a set of macroscopic parameters $\gamma_\alpha(t)$ characterizing the medium

$$\rho_1(t) = Sp_2 \rho(t); \quad \gamma_\alpha(t) = Sp\{\rho(t)\hat{\gamma}_\alpha\}. \tag{1}$$

Hereinafter $Sp_2 A$ means taking the trace of operator $A$ over medium variables. Linear independent operators $\hat{\gamma}_\alpha$ correspond to the parameters $\gamma_\alpha$ and are determined by the structure and symmetry properties of the Hamiltonian $H_2$ of the medium.

A simple method of construction of $\rho(t)$ consists in using the Liouville equation with the infinitely small source [7]

$$\left(\frac{\partial}{\partial t} + iL\right)\rho(t) = -\varepsilon\{\rho(t) - P(t)\rho(t)\}, \quad \varepsilon \to +0; \tag{2}$$

$$L = L_0 + L_V, \quad L_0 = L_1 + L_2, \quad iL_i A = \frac{1}{i\eta}[A, H], \quad iL_V A = \frac{1}{i\eta}[A, V], \quad j = 1,2.$$

The projection operator $P(t)$ in (2) determines the character of the reduced description of the non-equilibrium system states.

It is convenient to choose $P(t)$ as the projection operator of the Kawasaki-Gunton type [8]

$$P(t)A = \sigma_q(t) Sp_2 A + \rho_1(t) \sum_\alpha \frac{\partial \sigma_q(t)}{\partial \gamma_\alpha(t)} \{Sp(\hat{\gamma}_\alpha A) - \gamma_\alpha(t) Sp A\}, \tag{3}$$

where

$$\sigma_q(t) \equiv \sigma_q(\gamma(t)) = Q_q^{-1} \exp\left\{-\sum_\alpha F_\alpha(t)\hat{\gamma}_\alpha\right\}; \quad Q_q = Sp_2 \exp\left\{-\sum_\alpha F_\alpha(t)\hat{\gamma}_\alpha\right\} \tag{4}$$

is the quasi-equilibrium statistical operator of the medium dependent on the $t$ implicitly via macroscopic parameters $\gamma_\alpha(t)$. The quantities $F_\alpha(t) = F_\alpha(\gamma(t))$ are called generalized thermodynamic forces conjugated to parameters $\gamma_\alpha(t)$. The functional connection of $F_\alpha(t)$ with $\gamma_\alpha(t)$ is determined by the balancing conditions [1]:

$$Sp\rho(t)\hat{\gamma}_\alpha = Sp_2\sigma_q(t)\hat{\gamma}_\alpha. \tag{5}$$

From the explicit form of $P(t)$, it follows that

$$P(t)\rho(t) = \rho_1(t)\sigma_q(t); \quad P(t)\frac{\partial \rho(t)}{\partial t} = \frac{\partial}{\partial t}\{P(t)\rho(t)\}; \quad P(t')P(t)A = P(t')A. \tag{6}$$

Subtracting the expression $\left(\frac{\partial}{\partial t} + iL_0\right)P\rho$ from the both sides of (2) and using (6), we get

$$\left(\frac{\partial}{\partial t} + iL_0 + \varepsilon\right)(1-P)\rho = -iL_V\rho - iL_0P\rho - P\frac{\partial \rho}{\partial t}. \tag{7}$$

Let us act with the operator $P(t)$ on the equation (2) from the left:

$$P\frac{\partial \rho}{\partial t} = -PiL_0\rho - PiL_V\rho.$$

As a result, (7) takes the form:

$$\left(\frac{\partial}{\partial t} + iL_0 + \varepsilon\right)(1-P)\rho = -(1-P)iL_V\rho + PiL_0\rho - iL_0P\rho. \tag{8}$$

As soon as we are interested in times $t \gg \tau_0$, the following commutation relations should be fulfilled [2]:

$$iL_0\hat{\gamma}_\alpha \equiv iL_2\hat{\gamma}_\alpha = i\sum_\beta a_{\alpha\beta}\hat{\gamma}_\beta, \tag{9}$$

where $a_{\alpha\beta}$ is the $c$-numbers matrix determined by the Hamiltonian $H_2$. Taking into account Eq. (9) and formulae from [2],

$$\exp(-iL_0t')\sigma_q(\gamma(t)) = \sigma_q\{\exp(iat')\gamma(t)\}, \tag{10}$$

it is easy to show that

$$P(t)iL_0\rho(t) = P(t)iL_0P(t)\rho(t) = iL_0P(t)\rho(t).$$

To do this, it is sufficient to differentiate the both sides of (10) over $t'$ at the point $t' = 0$ and to use definition (3). As a result, Eq. (8) is simplified:

$$\left(\frac{\partial}{\partial t} + iL_0 + \varepsilon\right)(1-P)\rho = -(1-P)iL_V\rho. \tag{11}$$

This differential equation is equivalent to the integral equation for $\rho(t)$ [10]:

$$\rho(t) = P(t)\rho(t) - \int_{-\infty}^{0} dt' \exp\{(\varepsilon + iL_0)t'\}(1 - P(t+t'))iL_V \rho(t+t'). \tag{12}$$

Let us define the operator of the time shift by the relation

$$\exp\left(t'\frac{\partial}{\partial t}\right)f(t) = f(t+t').$$

Then equation (12) can be written in the form

$$\rho(t) = P(t)\rho(t) - \int_{-\infty}^{0} dt' \exp\{(\varepsilon + iL_0)t'\}\exp\left(t'\frac{\partial}{\partial t}\right)(1 - P(t))iL_V \rho(t)$$

and it has the formal solution

$$\rho(t) = \{1 + X(t)\}^{-1} P(t)\rho(t) \equiv \sum_{n=0}^{\infty} (-1)^n X^n(t) P(t)\rho(t), \tag{13}$$

where the operator $X(t)$ is given by the formula

$$X(t) = \int_{-\infty}^{0} dt' \exp\{(\varepsilon + iL_0)t'\}\exp\left(t'\frac{\partial}{\partial t}\right)(1 - P(t))iL_V .$$

It can be shown by the direct calculation that balancing conditions (5) are fulfilled automatically, while at $t \to -\infty$ the boundary condition of the coincidence of $\rho(t)$ and $P(t)\rho(t)$ is valid.

Due to the presence of the time shift operator, non-equilibrium statistical operator (13) depends on all parameter values $\rho_1(t')$, $\gamma_\alpha(t')$ prior to the moment $t' \leq t$. However, the "non-markovity" of $\rho(t)$ is in a certain sense fictitious, since, as it is shown below, by means of the rigorous transformation of the operator $\exp\left(t'\frac{\partial}{\partial t}\right)$ we succeeded to give $\rho(t)$ the Markovian form, which takes into account all memory effects contained in the usual non-Markovian recording of $\rho(t)$ [1].

The operators $P(t)$ and $P(t)\rho(t)$ depend on time via the values of the quantities $\rho_1(t)$, $\gamma_\alpha(t)$ taken at the same moment $t$. Therefore, the action of the operator $\exp\left(t'\frac{\partial}{\partial t}\right)$ in (13) is equivalent to the differentiation over these parameters

$$\exp\left(t'\frac{\partial}{\partial t}\right) \equiv \exp(t'D(t)); \quad D(t) = \sum_\alpha \frac{\partial \gamma_\alpha(t)}{\partial t}\frac{\partial}{\partial \gamma_\alpha(t)} + \frac{\partial \rho_1(t)}{\partial t}\frac{\partial}{\partial \rho_1(t)}.$$

Here the following notation is introduced

$$\frac{\partial \rho_1(t)}{\partial t}\frac{\partial}{\partial \rho_1(t)} \equiv \sum_{m,n} \frac{\partial \rho_{1mn}(t)}{\partial t}\frac{\partial}{\partial \rho_{1mn}(t)}.$$

The indices $m$ and $n$ number the arbitrary basis in the state space of the dynamic system. The parameter change rates entering $D(t)$ read as

$$\frac{\partial \rho_1(t)}{\partial t} = -Sp_2 iL\rho(t) \equiv \Lambda_1(t); \qquad \frac{\partial \gamma_\alpha(t)}{\partial t} = Sp\rho(t)iL\hat{\gamma}_\alpha \equiv \Lambda_\alpha(t). \qquad (14)$$

We obtain $\Lambda_1(t)$ by taking the trace of the both sides of Liouville equation (2) over the medium variables. Multiplication of (2) by $\hat{\gamma}_\alpha$ and the further calculation of the trace over all variables with the use of (4) gives $\Lambda_\alpha(t)$.

Suppose now that the time evolution of the non-equilibrium state is Markovian, i.e. the collision integrals $\Lambda_1(t)$, $\Lambda_\alpha(t)$, and, hence, also $D(t)$ are the functions of $\rho_1(t)$, $\gamma_\alpha(t)$. In such a case $X(t)$ can be written in the form

$$X(t) = \int_{-\infty}^{0} dt' \exp\{(\varepsilon + iL_0)t'\}\exp(t'D(\rho_1(t)\gamma(t)))(1-P(t))iL_V, \qquad (15)$$

$$D(\rho_1(t)\gamma(t)) = \sum_\alpha Sp\{\rho(t)iL\hat{\gamma}_\alpha\}\frac{\partial}{\partial \gamma_\alpha(t)} - Sp_2\{iL\rho(t)\}\frac{\partial}{\partial \rho_1(t)}.$$

Such a presentation of the operator $X(t)$ provides the Markovian character of $\rho(t)$ (13) which, in accordance with (14), agrees with the initial assumption on the memory absence in the collision integrals $\Lambda_1(t)$, $\Lambda_\alpha(t)$. So, there exist two completely equivalent forms of the solution of the Liouville equation with source (2): the Markovian one and the non-Markovian one[2].

Substituting (13) in (14), after simple calculations, we get the exact system of generalized kinetic equations of the Markovian type:

$$\frac{\partial \rho_1(t)}{\partial t} + iL_1\rho_1(t) = -Sp_2\{iL_V(1+X(t))^{-1}P(t)\rho(t)\},$$

$$\frac{\partial \gamma_\alpha(t)}{\partial t} - i\sum_\beta a_{\alpha\beta}\gamma_\beta(t) = Sp\{(1+X(t))^{-1}P(t)\rho(t)iL_V\hat{\gamma}_\alpha\}. \qquad (16)$$

Operator $(1+X(t))^{-1}$ is understood in the sense of its series expansion

$$(1+X(t))^{-1} \equiv \sum_{n=0}^{\infty}(-1)^n X^n(t),$$

while $X(t)$ is given by Eq. (15).

Let us adduce the explicit form of equations (16) in the second order of the perturbation theory over $V$. As it follows from the structure of the right sides of (16), it is sufficient to calculate the operator $(1+X(t))^{-1}P(t)\rho(t)$ in the first order, which gives

---

[2] It was pointed in Ref. [11] at the possibility of the construction of the non-equilibrium statistical operator in the form of the formal Markovian expansion in terms of a small interaction.

$$\frac{\partial \rho_1(t)}{\partial t} + iL_1\rho_1(t) \approx Sp_2\left\{iL_V(1-X(t))^{-1}P(t)\rho(t)\right\},$$

$$\frac{\partial \gamma_\alpha(t)}{\partial t} - i\sum_\beta a_{\alpha\beta}\gamma_\beta(t) \approx Sp\left\{(1-X(t))^{-1}P(t)\rho(t)iL_V\hat{\gamma}_\alpha\right\}, \quad (17)$$

where

$$X_1(t) = \int_{-\infty}^{0} dt' \exp\left\{(\varepsilon + iL_0)t'\right\}\exp(t'D_0(t))(1-P(t))iL_V,$$

$$D_0(t) = i\sum_{\alpha\beta} a_{\alpha\beta}\gamma_\beta(t)\frac{\partial}{\partial \gamma_\alpha(t)} - iL_1\rho_1(t)\frac{\partial}{\partial \rho_1(t)}.$$

Since, in the zero approximation over $V$, the relations

$$\exp(t'D_0(t))P(t)\rho(t) = \exp(-iL_0t')P(t)\rho(t),$$

$$\exp(t'D_0(t))P(t)iL_V P(t)\rho(t) = \exp(-iL_0t')P(t)iL_V(t')P(t)\rho(t),$$

$$iL_V(t')A \equiv \frac{1}{i\eta}\left[A, \exp(iL_0t')V\right]$$

are valid, we get

$$X_1(t)P(t)\rho(t) = (1-P(t))\int_{-\infty}^{0} dt' e^{\varepsilon t'} iL_V(t')P(t)\rho(t). \quad (18)$$

Using (18) and the explicit form (3) of the operator $P(t)$, it is not difficult to obtain the system of generalized kinetic equations in the second order of the perturbation theory over $V$:

$$\frac{\partial \rho_1(t)}{\partial t} = \frac{1}{i\eta}\left[\tilde{H}_1, \rho_1\right] -$$

$$-\frac{1}{\eta^2}\int_{-\infty}^{0} dt' e^{\varepsilon t'} Sp_2\left\{[[\rho_1\sigma_q, V(t')], \delta V] - \sum_\alpha \left[\rho_1, Sp_2\frac{\partial \sigma_q}{\partial \gamma_\alpha}V\right]Sp\rho_1\sigma_q[V(t'), \hat{\gamma}_\alpha]\right\}, \quad (19)$$

$$\frac{\partial \gamma_\alpha(t)}{\partial t} = \frac{1}{i\eta}Sp_2\sigma_q\left[\hat{\gamma}_\alpha\tilde{H}_2\right] -$$

$$-\frac{1}{\eta^2}\int_{-\infty}^{0} dt' e^{\varepsilon t'} Sp\left\{\sigma_q\rho_1[[\delta V(t')], [V, \hat{\gamma}_\alpha]] - \sum_\beta [V(t')\hat{\gamma}_\beta]Sp\rho_1\frac{\partial \sigma_q}{\partial \gamma_\beta}[V, \hat{\gamma}_\alpha]\right\}, \quad (20)$$

where

$$\tilde{H}_1 = H_1 + Sp_2(\sigma_q V), \quad \tilde{H}_2 = H_2 + Sp_1(\rho_1 V), \quad \delta V = V - Sp_2(\sigma_q V), \quad V(t') = \exp(iL_0t')V.$$

The quantities $Sp_1(\rho_1 V)$ and $Sp_2(\sigma_q V)$ play the role of self-consistent fields acting on the medium and the dynamic system due to the interaction between them. The structure of the second order terms over $V$ in (19) and (20) is such that the influence of the self-consistent fields is excluded in them. Eqs. (19) and (20) represent the closed system of the nonlinear kinetic equations for the variables $\rho_1(t)$, $\gamma_\alpha(t)$, which appropriately takes into account the inter-influence of the non-equilibrium medium and the dynamic system.

In the case when $V$ depends only on the medium coordinates, i.e. the interaction between the medium and the dynamic system is not taken into account, (20) transforms into the known equation from work [12], and (19) takes the form of the usual Liouville equation with the Hamiltonian $H_1$, as it must.

The kinetic equations of (19), (20) type can be derived by other method which is convenient in the case when the non-equilibrium behavior of the medium in the absence of the dynamic system is known and it is necessary to take into account only the interaction between the medium and the dynamic system. The corresponding derivation is given in the **Appendix**.

### 3. Evolution of the quantum system state in the medium

Let us consider the interaction of the dynamic system possessing a small number of degrees of freedom (an atomic or molecular system can serve as an example) with the non-equilibrium medium.

As is well known, the state of the isolated quantum system is described by the wave function satisfying the Schrödinger equation. If the system interacts with the medium, it ceases to be purely mechanical one, acquiring the statistical system pattern, which can be partially taken into account by ascribing the finite lifetimes to its stationary states [13]. For obtaining the explicit expressions of these times, they derive the Schrödinger-type equation with non-selfconjugate Hamiltonian, anti-Hermitian part of which determines the broadening of the system energy levels. However, this approach is approximate, since it does not reflect the fact that the system state in the medium is mixed and cannot be fully characterized by a definite new function, let it even be with the complex energy value. Of course, in a certain moment of time, the pure state can be "prepared" (for instance, by carrying out the complete system measurement), which afterwards turns into the mixed state under the influence of the medium.

In accordance with the above-mentioned, we shall seek the solution of Eq. (19), supposing that, at the initial time moment, the dynamic system is characterized by the wave-function $|\psi\rangle$ and, hence, by the statistical operator $\rho_1(0)|\psi\rangle\langle\psi|$.

Let $|n\rangle$ and $E_n$ be the eigen-function and the eigen-energy of Hamiltonian $H_1$ respectively,

$$H_1|n\rangle = E|n\rangle.$$

In the $H_1$ representation, Eq. (19) reads as follows

$$\frac{\partial \rho_{mn}}{\partial t} = -i\omega_{mn}\rho_{mn} + \frac{i}{\eta}\langle m|[\rho_1, Sp_2\sigma_q V]|n\rangle -$$

$$-\frac{1}{\eta^2}\int_{-\infty}^{0} dt' e^{\varepsilon t'} \sum_l \{\rho_{ml}\langle l|Sp_2\sigma_q V(t')\delta V|n\rangle + \rho_{ln}\langle m|Sp_2\sigma_q \delta V V(t')|l\rangle -$$

$$- Sp_2\sigma_q \sum_k \rho_{lk}(\langle k|\delta V|n\rangle\langle m|V(t')|l\rangle + \langle k|V(t')|n\rangle\langle m|\delta V|l\rangle) -$$

$$- \sum_\alpha \left(\rho_{ml} Sp_2 \frac{\partial \sigma_q}{\partial \gamma_\alpha} V_{ln} - \rho_{ln} Sp_2 \frac{\partial \sigma_q}{\partial \gamma_\alpha} V_{ml}\right) Sp\sigma_q \rho_1 [V(t'), \gamma_\alpha]\}, \quad (21)$$

$$\rho_{mn} \equiv \langle m|\rho_1|n\rangle, \quad \omega_{mn} = \frac{1}{\eta}(E_m - E_n).$$

In the $H_1$ representation, the initial condition of (21) is:

$$\rho_{mn}(0) = c_m(0)c_n^*(0), \quad |\psi\rangle = \sum_n c_n(0)|n\rangle \quad (22)$$

We seek the solution of (21) in the form:

$$\rho_{mn}(t) = c_m(t)c_n^*(t) + \rho_{mn}^c(t), \quad \rho_{mn}^c(0) = 0 \quad (23)$$

at the natural additional condition, according to which the coefficients $c_m(t)$ must satisfy the Schrödinger equation with the effective Hamiltonian. The substitution of (23) into (21) leads to the system of equations

$$i\eta \frac{\partial c_m}{\partial t} = E_m c_m + \sum_l Sp_2(\sigma_q V_{ml})c_l - \frac{i}{\eta}\sum_l c_l \int_{-\infty}^{0} dt' e^{\varepsilon t'} \langle m|Sp_2\sigma_q \delta V V(t')|l\rangle, \quad (24)$$

$$i\eta \frac{\partial \rho_{mn}^c}{\partial t} = \langle m|[H_1 + Sp_2\sigma_q V, \rho^c]|n\rangle - \frac{1}{\eta}\int_{-\infty}^{0} dt' e^{\varepsilon t'} \{\langle m|Sp_2[[\rho^c \sigma_q, V(t')], \delta V]|n\rangle -$$

$$- \sum_{l,k} c_l c_k^* Sp_2 \sigma_q (\delta V_{kn} V_{ml}(t') + V_{kn}(t')\delta V_{ml}) - \quad (25)$$

$$- \sum_{l\alpha}\left[(c_m c_l^* + \rho_{ml}^c)Sp_2 \frac{\partial \sigma_q}{\partial \gamma_\alpha} V_{ln} - (c_l c_n^* + \rho_{ln}^c)Sp_2 \frac{\partial \sigma_q}{\partial \gamma_\alpha} V_{ml}\right]\sum_{p,k} Sp_2\sigma_q(c_p c_k^* + \rho_{pk}^c)[V_{pk}(t')\hat{\gamma}_\alpha]\},$$

where $A_{kn} \equiv \langle k|A|n\rangle$. System (24), (25) is equivalent to equation (21). If we suppose that $\psi(t) = \sum_n c_n(t)|n\rangle$, (24) takes on the form of the Schrödinger-type equation

$$i\eta \frac{\partial \psi(t)}{\partial t} = (H_1 + Sp_2\sigma_q V)\psi(t) - \frac{i}{\eta}\int_{-\infty}^{0} dt' e^{\varepsilon t'} Sp_2\{\sigma_q \delta V V(t')\}\psi(t), \quad (26)$$

while (25) can be written in the form

$$i\eta\frac{\partial\rho^c}{\partial t}=\left[H_1+Sp_2\sigma_q V,\rho^c\right]-\frac{i}{\eta}\int_{-\infty}^{0}dt'e^{\varepsilon t'}\left\{Sp_2\left[\left[\rho^c\sigma_q,V(t')\right],\delta V\right]-Sp_2\sigma_q\left[V(t')P_{\psi(t)}\delta V+\right.\right.$$

$$\left.+\delta V P_{\psi(t)}V(t')\right]-\sum_\alpha\left[P_{\psi(t)}+\rho^c,Sp_2\frac{\partial\sigma_q}{\partial\gamma_\alpha}V\right]\sum_{p,k}Sp\sigma_q\left(P_{\psi(t)}+\rho^c\right)\left[V(t'),\gamma_\alpha\right]\right\}. \qquad (27)$$

Here $P_{\psi(t)}=|\psi(t)\rangle\langle\psi(t)|$ is the projection operator onto the state $\psi(t)$. According to (26), the effective Hamiltonian of the system in the medium is given by the expression

$$H_{\it eff}=H_1+Sp_2\sigma_q V-\frac{1}{\eta}\int_{-\infty}^{0}dt'e^{\varepsilon t'}Sp_2\{\sigma_q\delta V\ V(t')\}. \qquad (28)$$

In the general case of the non-equilibrium medium, $H_{\it eff}$ depends on time through parameters $\gamma_\alpha(t)$. If the medium is either in equilibrium or in a stationary non-equilibrium state, the effective Hamiltonian $H_{\it eff}$ does not depend on time. We denote

$$K=-\frac{1}{\eta}\int_{-\infty}^{0}dt'e^{\varepsilon t'}Sp_2\{\sigma_q\delta V\ V(t')\}\equiv U-i\Gamma,$$

where

$$U=\frac{K+K^+}{2},\quad \Gamma=\frac{K^+-K}{2i}$$

are the Hermitian and anti-Hermitian part of the non-self-conjugate operator $K$, respectively. It is obvious that

$$U=-\frac{1}{2i\eta}\int_{-\infty}^{0}dt'e^{\varepsilon t'}Sp_2\{\sigma_q[\delta V,V(t')]\},\ \Gamma=\frac{1}{2\eta}\int_{-\infty}^{0}dt'e^{\varepsilon t'}Sp_2\{\sigma_q(\delta V\ V(t')+V(t')\delta V)\}, \qquad (29)$$

$$H_{\it eff}=H_1+\delta H_1-i\Gamma,\ \delta H_1=Sp_2\sigma_q V+U. \qquad (30)$$

The Hermitian operators $\delta H_1$ and the $\Gamma$ condition denote the shift of energy levels and the decay of the eigen-states of the Hamiltonian $H_1$, respectively. Therefore, it is natural to call them the shift and decay operators of the system. In the general case, $\delta H_1$ and $\Gamma$ depend on time. So, the evolution of the dynamic system interacting with the medium is described by the statistical operator $\rho_1=P_{\psi(t)}+\rho^c(t)$, where $\psi(t)$ and $\rho^c(t)$ satisfy equations (26), (27) and initial conditions (22), (23). Under the influence of the medium, the initial pure state of the system $\rho_1\approx P_{\psi(t)}$ decays with the characteristic time $\Gamma^{-1}$ ($P_{\psi(t)}\to 0$ at $t\gg\Gamma^{-1}$), which can be considered as the time of the system transition into the mixed state ($\rho_1\to\rho^c$). It is obvious from the above-mentioned that the approach based on the effective Schrödinger equation for the system in the medium is applicable at times $t<\Gamma^{-1}$, while for times $t\gg\Gamma^{-1}$, when $\rho_1\approx\rho^c$, the evolution is described by equation (27), which at such times coincides with (19). In a particular case of the particle interaction with the equilibrium medium, (26) coincides with the equation obtained in paper [14].

## 4. Kinetic equation for impurity particles in a liquid

As a next example of application of system (19), (20), we will consider the derivation of the kinetic equation for spinless particles in the equilibrium liquid. At first we shall restrict ourselves by a homogeneous case when the partition function $f(p,t)$ of impurity particles does not depend on coordinates and is defined by the formula [2]:

$$f(p,t) = Sp_1\{\rho_1(t) a_p^+ a_p^-\},$$

where $a_p^+$ and $a_p^-$ are the creation and annihilation operators of the impurity particle with the momentum $p$, respectively. Multiplying the both parts of Eq. (21) by $a_p^+ a_p^-$ and taking the trace over the states of the dynamic system, we get

$$\frac{\partial f(p,t)}{\partial t} = \frac{i}{\eta} Sp \rho_1 \sigma [H_1 + V, a_p^+ a_p^-] -$$

$$-\frac{1}{\eta^2} \int_{-\infty}^{0} dt' e^{\varepsilon t'} \left\{ Sp \rho_1 \sigma [[V(t')],[V, a_p^+ a_p^-]] - \sum_\alpha Sp \rho_1 \sigma [V(t') \hat{\gamma}_\alpha] \frac{\partial}{\partial \gamma_\alpha} Sp \rho_1 \sigma [V, a_p^+ a_p^-] \right\}. \quad (31)$$

The coarse statistical operator $\sigma$ of the equilibrium liquid is given by Gibbs's great canonical distribution:

$$\sigma = \exp\{\Omega - \beta(H_2 - \mu \hat{N})\}, \quad (32)$$

where $\Omega$ and $\mu$ are the thermodynamic and chemical potentials of liquid, respectively; $\beta$ is the inverse temperature, $\hat{N}$ is the operator of the total number of liquid particles.

We write the interaction $V$ between the particle and the liquid in the form:

$$V = \int dr dr' \psi^+(r) \psi(r) u(r-r') \varphi^+(r') \varphi(r') .$$

Operators $\varphi^+$ and $\varphi$ describe the creation and the annihilation of liquid particles; $u$ is the interaction potential; $\psi^+$ and $\psi$ are connected with $a_p^+$ and $a_p^-$ by means of the usual formulae

$$\psi(\vec{r}) = v^{-1/2} \sum_p a_p^- \exp\left(\frac{i}{\eta} \vec{p} \vec{r}\right), \quad \psi^+(\vec{r}) = v^{-1/2} \sum_p a_p^+ \exp\left(-\frac{i}{\eta} \vec{p} \vec{r}\right). \quad (33)$$

Here $v$ is the system volume. At low impurity concentration it is possible to neglect the impurity-impurity interaction and to consider $H_1$ as free particle Hamiltonian $H_1 = \sum_p \varepsilon_p^- a_p^+ a_p^-$, where $\varepsilon_p^- = p^2/2M$ is the energy of the impurity particle with the momentum $p$ and the mass $M$. In the approximation linear over impurity concentration, the equation for $f(p,t)$ does not depend on the impurity particle statistics. Besides, as to this approximation, it is possible to neglect the second term in the integrand of (31). Because of the liquid density constancy and of the relation $[H_1, a_p^+ a_p^-] = 0$, the contribution of the first term of the right-hand side of (33) turns to zero. After the above-mentioned simplifications and with account for (32), Eq. (33) takes the form:

$$\frac{\partial f_p^-}{\partial t} = -\frac{1}{\eta^2} \int_{-\infty}^{0} dt' e^{\varepsilon t'} Sp\{\rho_1 \sigma [V(t'), [\delta V, a_p^+ a_p^-]]\}. \tag{34}$$

Calculating the commutators entering (34), making the usual uncouplings of the type

$$\langle a_1^+ a_2^+ a_3 a_4 \rangle = \langle a_1^+ a_3 \rangle \langle a_2^+ a_4 \rangle + \langle a_2^+ a_3 \rangle \langle a_1^+ a_4 \rangle$$

and leaving only linear terms over $f(p,t)$, after straightforward transformations, we get the kinetic equation in the form

$$\frac{\partial f(p,t')}{\partial t} = -v^{-1} \sum_{p'} \left( W_{p'p} f_{p'}^- - W_{pp'} f_p^- \right). \tag{35}$$

Equation (35) has the form of the Pauli equation, the transition probabilities $W_{pp'}$ being connected with the scattering of the impurity particle on the liquid density fluctuations:

$$W_{pp'} = \int_{-\infty}^{\infty} dt' \int d\vec{r} d\vec{r}\,' d\vec{x}\, u(\vec{r}) u(\vec{r}\,') \times$$

$$\times \exp\left\{-\frac{i}{\eta}(\vec{p}-\vec{p}\,')(\vec{r}\,'-\vec{r}+\vec{x})\right\} \exp\left\{\frac{i}{\eta}(\varepsilon_p^- - \varepsilon_{p'}^-)t'\right\} \langle \delta N(\vec{x},t') \delta N(0,0) \rangle, \tag{36}$$

where the correlator of the liquid density fluctuations has the form:

$$\langle \delta N(r,t) \delta N(r',0) \rangle = Sp_2 \sigma \{\varphi^+(r,t)\varphi(r,t) - \langle \varphi^+ \varphi \rangle\}\{\varphi^+(r')\varphi(r') - \langle \varphi^+ \varphi \rangle\},$$

$$\langle \varphi^+ \varphi \rangle = Sp_2 \sigma\, \varphi^+(r,t)\varphi(r,t).$$

In the case of weak inhomogeneity, it is possible by a similar method to obtain the following equation for the distribution function

$$\frac{\partial f_p^-(r,t)}{\partial t} + \frac{\partial \varepsilon_p^-}{\partial p} \frac{\partial f_p^-(r,t)}{\partial r} = -v^{-1} \sum_{p'} \left( W_{p'p} f_{p'}^-(\vec{r},t) - W_{pp'} f_p^-(\vec{r},t) \right). \tag{37}$$

The probabilities $W_{pp'}$ are given by expressions (36), as before.

For the heavy impurity particle when $m/M \ll 1$ ($m$ is the mass of the liquid particle), the integro-differential equation (37) comes to a differential one. Really, in this case the relative change of the impurity particle momentum takes place $(\Delta p/p)\sqrt{m/M} \ll 1$. Hence it follows that $W_{pp'}$ has a sharp maximum at $p' \approx p$. Expanding $f_{p'}(r,t)$ in series in terms of $p' - p$ and accounting for relation $\sum_{p'} W_{p'p} = \sum_{p'} W_{pp'}$, we get:

$$\frac{\partial f_{\vec{p}}}{\partial t} + \frac{\partial \varepsilon_{\vec{p}}}{\partial p} \frac{\partial f_{\vec{p}}}{\partial r} = \frac{\partial J_{\vec{p}}}{\partial p},$$

$$\vec{J}_{\vec{p}} = v^{-1} \sum_{\vec{p}'} (\vec{p} - \vec{p}') W_{\vec{p}\vec{p}'} f_{\vec{p}} - (2v)^{-1} \sum_{\vec{p}',\beta} (\vec{p} - \vec{p}')(\vec{p} - \vec{p}')_\beta \frac{\partial f_{\vec{p}}}{\partial p_\beta} W_{\vec{p}\vec{p}'}. \tag{38}$$

It follows from the symmetry considerations that

$$v^{-1} \sum_{\vec{p}'} (p' - p) W_{\vec{p}\vec{p}'} = p\varphi(p),$$

$$v^{-1} \sum_{\vec{p}'} (\vec{p} - \vec{p}')_\alpha (\vec{p} - \vec{p}')_\beta W_{\vec{p}\vec{p}'} = c(\vec{p})\delta_{\alpha\beta} + b(p)\frac{p_\alpha p_\beta}{p^2}. \tag{39}$$

Taking into considerations Eq. (39), Eq. (38) takes on the known form of the Fokker-Planck equation for the Brownian particle distribution function [15], if the quantities $\varphi$ and $c$ in formulae (39) are assumed to be momentum-independent and $b(p) = 0$ is taken.

However, in the given case the coefficients obtained with the help of formulae (36) and (39) are expressed explicitly through the scattering of the impurities on the liquid density fluctuations.

### 5. Equation for the statistical operator of the spin-system in a thermostat

As the last example, we shall consider the derivation of the kinetic equation the statistical operator $\rho_1$ of a spin-system interacting with the equilibrium thermostat (lattice). The Hamiltonian of the system reads

$$H = H_S + H_L + H_{SL}.$$

Here $H_S$ and $H_L$ are the Hamiltonians of the spin-system and the lattice, respectively. $H_{SL}$ is the spin-lattice interaction. In the given case, we have $H_1 = H_S$, $H_2 = H_L$, $V = H_{SL}$. Since the lattice heat capacity is much greater than that of the spin-system, we can assume that the lattice is in the equilibrium state at the constant inverse temperature, and it is described by the canonical Gibbs distribution $\sigma_0 = \exp(-\beta_L H_L) / Sp_L \exp(-\beta_L H_L)$.

Hence, the medium is characterized by one macroscopic parameter $\gamma_\alpha$, which can be chosen as the lattice average energy $E_L = Sp_L \sigma_0 H_L$, so that $\hat{\gamma}_\alpha = H_L$. Inasmuch $[\sigma_0, H_L] = 0$, Eq. (19) takes the form:

$$\frac{\partial \rho_1(t)}{\partial t} = \frac{i}{\eta}[\rho_1, H_S + Sp_L \sigma_0 H_{SL}] -$$

$$-\frac{1}{\eta^2}\int_{-\infty}^{0}dt' e^{\varepsilon t'} Sp_L \left[\left[\rho_1\sigma_0, H_{SL}(t')\right], H_{SL} - Sp_L\sigma_0 H_{SL}\right]. \quad (40)$$

If $Sp_L\sigma_0 H_{SL} = 0$, Eq. (40) coincides with the well-known Bloch-Wangsness-Redfield equation [16], with the difference that, instead of $\rho_1 - \sigma_0$, the product $\rho_1\sigma_0$ enters Eq. (40), there is "cutting off" factor $e^{\varepsilon t'}$, and the integration over $t'$ from the very beginning is carried with the infinite lower boundary. We emphasize, however, that we derive Eq. (40) consequently without additional assumptions apart from those that constitute the basis of the modern theory of the reduced description of the non-equilibrium processes. In particular, it was not required that the statistical operator $\rho$ of the complete system (spin + lattice) had the multiplicativity property $\rho = \rho_1\sigma_0$ in all time moments. In the given consideration, this condition is applied only in the remote past $t \to -\infty$.

### 6. Conclusion

The method stated in the given paper is applicable to a lot of irreversible processes. In particular, it allows formulating the consistent theory of magnetic resonance in liquids, the theory of molecular motion of different complexes in solids, the theory of irreversible processes in coupled spin-phonon systems, etc. The subsystem selection in all these problems is not difficult. In addition to the problems of the above-mentioned type, with the help of exact kinetic equations (16), it is possible to investigate systems with strong fluctuations [17] if strongly fluctuating freedom degrees are considered as a dynamic system and described by their statistical operator $\rho_1(t)$, while the remaining freedom degrees are characterized by macroscopic parameters $\gamma_\alpha(t)$. The systems with the "fast" and "slow" freedom degrees can be also included into this scheme. The use of the statistical operator $\rho_1(t)$ allows describing the irreversible processes in classical as well as in quantum systems in the same manner. Note that $\rho_1(t)$ replaces the distribution function $f(a_1,...a_n, t)$ of the „crude" variables $a_1,...a_n$, which is used in paper [17] at the derivation of the Fokker-Planck equation for the classical and quantum systems.

The main difficulty in the mentioned situations lies in the physically justified division of the freedom degrees into groups and in the obtaining of the adequate expression for the interaction $V$ between these groups. This question needs an independent investigation for each particular problem.

In conclusion the authors express their deep gratitude to D.N. Zubarev for the valuable remarks and to V.G. Bar'yakhtar, L.L. Buishvili, and V.P. Kalashnikov for the useful discussions.

### Appendix

The derivation of the generalized kinetic equations of the Markovian type differing from that brought in the main text is given in the present Appendix. Let us proceed from the following separation of the system Hamiltonian:

$$H = H_1 + H_2 + V,$$

in which $H_2$ is the Hamiltonian of the medium including all its internal interactions, and $V$ is the interaction of the dynamic system with the medium. Denote the coarse statistical operator of the medium in the absence of the dynamic system by $\sigma(t) \equiv \sigma(\gamma_\alpha^0(t))$, where $\sigma(t)$ is the solution of the Liouville equation with the infinitely small source

$$\frac{\partial \sigma(t)}{\partial t} + iL_2 \sigma(t) = -\eta \{\sigma(t) - \sigma_q(t)\}, \quad \eta \to +0 \tag{A1}$$

depending on time implicitly, through the „undisturbed" macroscopic parameters

$$\gamma_\alpha^0(t) = Sp_2 \sigma(t) \hat{\gamma}_\alpha = Sp_2 \sigma_q(t) \hat{\gamma}_\alpha. \tag{A2}$$

Quantity $\sigma_q(t)$ is quasi-equilibrium medium operator (4) depending on $\gamma_\alpha^0(t)$; $\sigma(t)$ describes the non-equilibrium medium at $V = 0$; the explicit form of the function $\sigma(\gamma_\alpha^0(t))$ is supposed to be known.

For our purposes it is more suitable to express the solutions of Liouville equation as $\sigma(\gamma_\alpha^0(t))$ and $\rho_1(t)$, and later, after establishing the relation between $\gamma_\alpha^0(t)$ and $\gamma_\alpha(t)$, to find the function $\rho(\gamma_\alpha^0(t), \rho_1(t))$. Such a state of the problem on finding the non-equilibrium statistical operator corresponds to the quantum-mechanical perturbation theory, when the solution of the perturbed problem is found with the help of that of unperturbed problem which is supposed to be known.

From the above reasoning, we choose the projection operator in the form

$$P(t)A = \sigma(t) Sp_2 A. \tag{A3}$$

The choice (A3) maximally uses the information on the medium in the absence of the dynamic system.

It is not difficult to make sure that the projection operator possesses the following properties:

$$\tilde{P}(t)\rho(t) = \rho_1(t)\sigma(t), \quad \tilde{P}(t)\frac{\partial \rho}{\partial t} = \sigma(t)\frac{\partial \rho_1(t)}{\partial t}, \quad \tilde{P}(t')\tilde{P}(t)A = \tilde{P}(t')A. \tag{A4}$$

Subtracting from the both sides of the Liouville equation

$$\left(\frac{\partial}{\partial t} + iL\right)\rho = -\varepsilon(1-\tilde{P})\rho \tag{A5}$$

the expression $\left(\frac{\partial}{\partial t} + iL_0\right)\tilde{P}\rho$, we get

$$\left(\frac{\partial}{\partial t} + iL_0 + \varepsilon\right)(1-\tilde{P})\rho = -iL_V\rho - iL_0\tilde{P}\rho - \frac{\partial}{\partial t}\tilde{P}\rho. \tag{A6}$$

The use of Eqs. (A1, A3, A4) gives:

$$\frac{\partial}{\partial t}\widetilde{P}\rho = \frac{\partial}{\partial t}(\sigma\rho_1) = \widetilde{P}(t)\frac{\partial\rho}{\partial t} - iL_2\widetilde{P}\rho. \tag{A7}$$

Acting by the operator $\widetilde{P}$ on the equation (A5) from the left, we obtain

$$\widetilde{P}\frac{\partial\rho}{\partial t} = -i\widetilde{P}(L_0 + L_V)\rho = -iL_1\widetilde{P}\rho - \widetilde{P}iL_V\rho, \tag{A8}$$

inasmuch $\widetilde{P}iL_2\rho = 0$, $\widetilde{P}iL_1\rho = iL_1\widetilde{P}\rho$. Therefore (A7) takes the form

$$\frac{\partial}{\partial t}\widetilde{P}\rho = -iL_0\widetilde{P}\rho - \widetilde{P}iL_v\rho. \tag{A9}$$

The substitution of Eq. (A9) into (A6) leads again to differential equation (11) and, correspondingly, to integral equation (12)

$$\rho(t) = \widetilde{P}(t)\rho(t) - \int_{-\infty}^{0} dt' e^{\varepsilon t'} e^{iL_0 t'}\left(1 - \widetilde{P}(t+t')\right)iL_V\rho(t+t'), \tag{A10}$$

but with the other projection operator $\widetilde{P}(t)$ (A3).

With the help of obvious relations

$$e^{iL_0 t'}\sigma(t+t') = \sigma(t), \quad e^{iL_0 t'}P(t+t') = \widetilde{P}(t), \quad \rho(t+t') = e^{-iLt'}\rho(t)$$

Eq. (A10) can be written in the form

$$\rho(t) = \widetilde{P}(t)\rho(t) - \left(1 - \widetilde{P}(t)\right)\int_{-\infty}^{0} dt' e^{\varepsilon t'} iL_V(t')e^{iL_0 t'}e^{-iLt'}\rho(t)$$

$$iL_V(t')A = \frac{1}{i\eta}[A, V(t')] \tag{A11}$$

and it has the formal solution

$$\rho(t) = \{1 + \widetilde{X}(t)\}^{-1}\widetilde{P}(t)\rho(t), \tag{A12}$$

where

$$\widetilde{X}(t) = \left(1 - \widetilde{P}(t)\right)\int_{-\infty}^{0} dt' e^{\varepsilon t'} iL_V(t')e^{iL_0 t'}e^{-iLt'}. \tag{A13}$$

According to the general ideology, $\rho(t)$ is the function of $\rho_1(t)$ and $\gamma_\alpha(t)$, while Eq. (A12) is expressed $\rho(t)$ as $\rho_1(t)$ and the auxiliary quantities $\gamma_\alpha^0(t)$.

With the help of (1) and (A12), it is easy to express $\gamma_\alpha^0(t)$ as $\gamma_\alpha(t)$:

$$\gamma_\alpha^0(t) = \gamma_\alpha(t) + Sp\left\{\hat{\gamma}_\alpha\left(1 + \widetilde{X}(t)\right)^{-1}\widetilde{X}(t)\widetilde{P}(t)\rho(t)\right\}. \tag{A14}$$

Similar to the method of derivation of Eqs. (16), the system of the kinetic equations of the Markovian type can be obtained:

$$\frac{\partial \rho_1}{\partial t} + iL_1 \rho_1 = -Sp_2 \left\{ iL_V \left(1 + \widetilde{X}(t)\right)^{-1} \widetilde{P}(t) \rho(t) \right\},$$

$$\frac{\partial \gamma_\alpha(t)}{\partial t} = Sp \left\{ \left(1 + \widetilde{X}(t)\right)^{-1} \widetilde{P}(t) \rho(t) iL \hat{\gamma}_\alpha \right\}. \tag{A15}$$

Equations (A15) can be considered as the parametric form of the generalized kinetic equations [18].

Excluding $\gamma_\alpha^0(t)$ with the help of Eq. (A14), it is possible, in principle, to obtain the closed system of the kinetic equations for $\rho_1(t)$ and $\gamma_\alpha(t)$. The structure of the system (A15) and of the relation (A14) is such that it allows us to get easily the arbitrary approximation over $V$.